\documentclass[%
 reprint,
 amsmath,amssymb,
 aps,
]{revtex4-2}
\usepackage{graphicx}
\usepackage{dcolumn}
\usepackage{bm}
\usepackage{xcolor}
\usepackage{hyperref}
\usepackage[normalem]{ulem}
\usepackage{float}
\usepackage{soul}

\preprint{APS/123-QED}

\begin{document}
\title{Dynamical phase transitions in postictal generalized EEG suppression}
\author{Subhasanket Dutta$^1$}
\email{subhosanket@gmail.com}
\author{Sarika Jalan$^1$}\email{Corresponding Author sarika@iiti.ac.in}
\author{Yash Shashank Vakilna$^2$}
\email{yashvakilna@gmail.com}
\author{Sandipan Pati$^2$}
\email{sandip@umn.edu}
\affiliation{$^1$ Complex Systems Lab, Department of Physics, Indian Institute of Technology Indore, Khandwa Road, Simrol, Indore-453552, India}
\affiliation{$^2$ Texas Institute for Restorative Neurotechnologies, The University of Texas Health Science Center at Houston, Houston, TX-77225, USA}

\date{\today}

\begin{abstract}
{
Postictal generalized EEG suppression (PGES) is a neurological condition that occurs in patients with generalized tonic-clonic seizures. It is marked by suppressed signals just after the seizure before the brain gradually recovers. Recovery from PGES involves a mixed state of amplitude suppression and high-amplitude oscillations, exhibiting a bimodal exponential distribution in power, unlike the unimodal exponential distribution of PGES. In this study, using the subcritical Hopf model, we explain the nature of phase transitions that underlie PGES. Our results reveal that recovery from PGES involves a change from a fixed point state to a bistable state (mixed phase), effectively captured by the noisy fixed-point and bistable regimes of the model. Consistent patterns across patients suggest a universal dynamical signature in PGES recovery. Our findings offer a mechanistic understanding of seizure termination and postictal brain state transitions.
}
\end{abstract}

\pacs{89.75.Hc, 02.10.Yn, 5.40.-a}

\maketitle
 \paragraph{{\bf Introduction:}}
 Abnormal oscillatory activity in the brain and other physiological systems is frequently linked to pathological conditions arising from disruptions in underlying neurological control mechanisms \cite{milton1989complex}. To gain deeper insights into these phenomena, a range of mathematical models have been developed, offering a framework to investigate how variations in system parameters can lead to deviations from normal function. Such deviations often manifest as bifurcations or phase transitions, providing a mechanistic understanding of the onset and progression of disease states \cite{ashwin2016mathematical}.
 
Such pathological dynamics include the emergence of hypersynchronous oscillations during epileptic seizures \cite{ren2021connectivity, truccolo2014neuronal}, or the loss of diurnal cortisol oscillations—typically modeled as a shift to a stable equilibrium—in patients with depression \cite{hollister1980hormones}. Altered neuronal synchrony across temporal and spatial scales has been implicated in several brain disorders, including Alzheimer’s disease, Parkinson’s disease, and schizophrenia \cite{uhlhaas2006neural, aron2016neural}. Furthermore, the spread of focal seizures has been modeled as a phase transition in computational frameworks of epilepsy \cite{moosavi2022critical}, reinforcing the relevance of bifurcation dynamics in pathological brain states.

One such condition is postictal generalized EEG suppression (PGES), a transient state that occurs in most patients following a generalized tonic-clonic seizure and is strongly associated with an increased risk of sudden unexpected death in epilepsy (SUDEP). PGES is characterized by a generalized suppression of EEG activity to amplitudes below $10 \mu V$ (in scalp EEG recordings) within 30 seconds after the seizure, ignoring the presence of muscle movements, breathing, and electrode artifacts \cite{lhatoo2010electroclinical, surges2011postictal, rajakulendran2015postictal, xu2016postictal}. The duration of PGES has emerged as a potential clinical biomarker, as prolonged suppression is correlated with heightened SUDEP risk \cite{marchi2019postictal, tao2013tonic, mier2020categorisation}.
\begin{figure*}[t!]    
\includegraphics[width=0.99\linewidth]{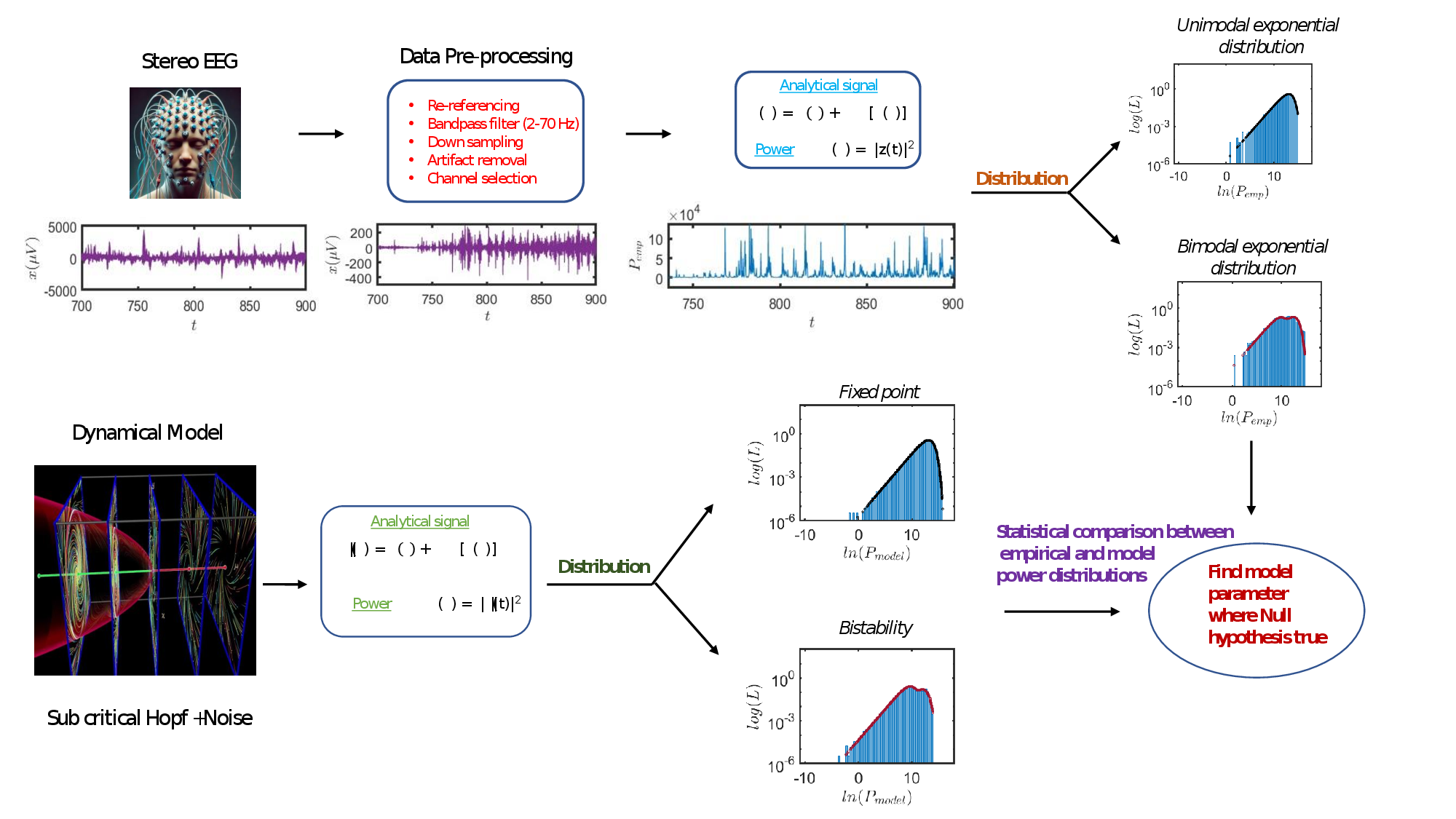}
  \caption{Flowchart illustrating the sequence of techniques used in this Letter, starting from EEG data and the dynamical model to the determination of model parameters where the model and empirical data align.}
  \label{fig:flowchart}
\end{figure*} 
In neuroscience, various efforts have been made to model neurological conditions as dynamical systems. A common approach involves comparing functional connectivity patterns derived from empirical data with those generated by computational models \cite{liu2023task, cabral2012modeling}. The time evolution of functional connectivity has been analyzed using methods such as the multiplication of temporal derivatives and the construction of functional connectivity dynamics matrices.
 \cite{shine2015estimation, hutchison2013dynamic}. Dynamical models have been particularly useful in exploring the resting state of the brain, where complex patterns of neural activity emerge in the absence of explicit stimuli \cite{deco2012ongoing, cabral2017functional, cabral2012modeling}. It has been proposed that multistability in these models can account for the temporal fluctuations and spatial patterns observed in resting-state signals \cite{deco2012ongoing, freyer2009bistability}. In particular, noise-induced switching between coexisting dynamical states has been shown to effectively reproduce the variability found in empirical recordings \cite{deco2017single, cabral2017functional}. Ghosh et al. further hypothesized that certain brain regions may operate near the critical point of a supercritical Hopf bifurcation, where fluctuations in coupling strength can drive transitions between a fixed point and a limit cycle, thereby capturing the dynamic range of brain activity observed in neurophysiological data \cite{ghosh2008cortical}.
Taking a different approach, Freyer {\it et al.} utilized a stochastic Hopf model to compare power distributions with their empirical counterparts, rather than focusing solely on functional connectivity \cite{freyer2009bistability,freyer2012canonical}, ultimately concluding that multi-stability could be a key factor in resting-state brain signals. In this study, we adopt a similar approach to investigate these dynamical properties further.

\begin{figure*}
 \includegraphics[width=0.99\linewidth]{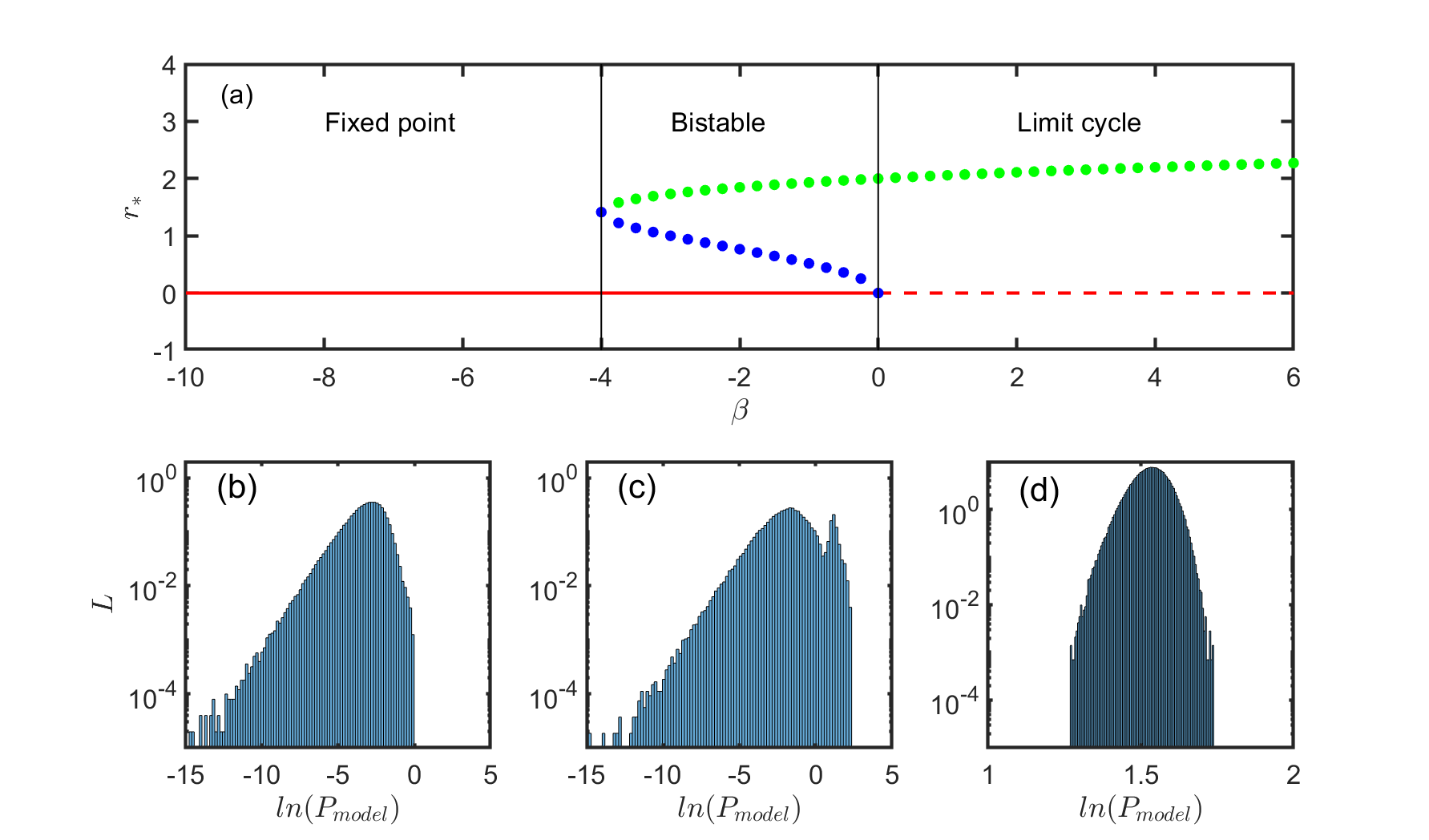}\\
 \caption{(a) The bifurcation plot $r^*$ vs.  $\beta$ for Eq.~ \ref{eqn:sub_hopf} at $\lambda=4$ and $\omega=3.0$. The red solid and dashed line corresponds to stable fixed point, and unstable fixed point, respectively. Blue and green circles represent the unstable and stable limit cycle, respectively.   (b), (c), (d) depict the power distribution at $\beta=-8$ (fixed point region), $\beta=-3.0$ (bistable region) and at $\beta=1.0$ (limit cycle region).}    \label{fig:bif} 
\end{figure*}
Although bifurcations at the onset and offset of epileptic seizures have been extensively studied \cite{breakspear2006unifying,jirsa2014nature,nazarimehr2018does}, to our knowledge, there exists no study on the phase transitions during the return to normal brain function. This Letter investigates the mechanisms behind postictal generalized PGES and the revival of normal brain function after PGES. We provide evidence that in the $\delta$ band, PGES primarily consists of a suppressed low-power state, which is followed by a bistable region consisting of switching between high- and low-power states before returning to the normal state. Suppression of oscillations is a well-known dynamical phenomenon that can be achieved by tuning the model parameters. The normal form of the Hopf bifurcation model provides a prototypical framework for such systems, where oscillations cease by adjusting the model parameters for a single oscillator. Additionally, in coupled oscillators with various coupling forms, such as conjugate coupling \cite{verma, dutta1}, different mechanisms of oscillation suppression can arise, including amplitude death and oscillation death.
This Letter reports that the transition from PGES to the normal state can be modeled using the normal form of a subcritical Hopf bifurcation, and it identifies the corresponding parameters of the differential equation that best replicate the empirical data in different states. We summarize the structure of this study in the form of a flow chart in Fig.~\ref{fig:flowchart}.
\\

\paragraph{\bf {Methods and techniques:}}

\subparagraph{{Dataset.}}
The data set is a stereo EEG time series of 5 subjects with one seizure each, and 1 subject with 2 seizures containing 276 channels with a sampling frequency of 2048 Hz. The data set was first referenced to a bipolar montage, following which it was filtered using the Finite impulse response (FIR) bandpass with a frequency range of $2Hz-70Hz$. Subsequently, the line noise was removed, and a notch filter was applied at $60Hz$. Afterwards, nine-time series were selected to represent the following areas - Anterior Hippocampus, Posterior hippocampus, Orbitofrontal, Cingulate, Frontal, Temporal, Insula, Amygdala, and Thalamus. The data was clipped to 10 minutes before and after a seizure. The duration of the ictal state differs from subject to subject. For an analytical EEG time series, the power at a particular time is defined as the square of the amplitude at that instant. The initial data set is a real-valued time series ($x(t)$). The Hilbert transform of $x(t)$ is given by $$\bar{x(t)}=\frac{1}{\pi} PV \int_{-\infty}^{+\infty}\frac{x(\tau)}{t-\tau}.$$ PV represents the Cauchy principal value. The analytical signal is defined as $x(t)+i\bar{x(t)}$, and hence the amplitude of the signal at time $t$ can be described as $\sqrt{x(t)^2+\bar{x(t)^2}}$.
 Moreover, to get the power time series for a particular frequency band, the real-valued signal should be passed through a bandpass filter of the desired bandwidth (for example, $2-4 Hz$ for $\delta$ band) before calculating the analytical signal. Another way to get a similar outcome is to use the Morlet wavelet transform \cite{cohen_book}.
 
\subparagraph{{Exponential distribution. }}
The central limit theorem states that the average of a large number of identical and independent random events tends to follow a Gaussian distribution. This theorem holds even when the random variables are not Gaussian. Applying a similar principle, one could argue that EEG signals may exhibit Gaussian characteristics \cite{feller}. Each electrode in a scalp EEG captures millions of neurons firing independently in a temporally uncorrelated manner, not necessarily following a Gaussian distribution. However, the central limit theorem suggests that the combined effect measured at an electrode should approximate a Gaussian distribution. Any deviation from normality would indicate a violation of the basic assumptions of the theorem. Based on this, it was proposed that EEG signals may generally be Gaussian processes \cite{gonen2012techniques,lion1953method,saunders1963amplitude}. However, later studies suggested that these results may depend on the length of the signal segment considered for the distribution \cite{mcewen1975modeling}. Longer signal segments were concluded to reduce the likelihood of obtaining a normal distribution. In addition to segmentation, other factors, such as sampling frequency and patient states, also affect normality \cite{gonen2012techniques}. A comparison between the amplitude distributions of task-dependent states and the resting state revealed that the former were less likely to follow a Gaussian distribution.

To prove the hypothesis, let us assume that the real-valued signal is a random variable $X$, and its Hilbert transform is another random variable $Y$. Both of these have Gaussian fluctuations, and their marginal distribution can be represented by a Gaussian distribution with zero mean and equal variance ($\sigma$) by $ f_X(x)=\frac{1}{\sigma\sqrt{2\pi}} exp(\frac{-x^2}{2\sigma})$ and $ f_Y(y)=\frac{1}{\sigma\sqrt{2\pi}} exp(\frac{-y^2}{2\sigma})$, respectively. Next, since $X$ and $Y$ are orthogonal to each other (as  $Y$ is Hilbert transform of $X$), the covariance matrix is given by, 
$
\begin{pmatrix}
\sigma^2 & 0 \\
0 & \sigma^2.
\end{pmatrix}
$
Therefore, the joint probability distribution is defined as $f_{X,Y}(x,y)= \frac{1}{\sigma\sqrt{2\pi}} exp(\frac{-(x^2+y^2)}{2\sigma})$.
 Since the amplitude is defined as $R=\sqrt{X^2+Y^2}$ and the power as $P=R^2$, we perform a variable transformation of $P=X^2+Y^2$ in the joint probability distribution to obtain the distribution of the power. The general equation for the transformation of variables in a probability distribution function is given by $P_Y(y)= \left|J\right|P_X(x)$ and we obtain $P_X(x)=\eta \: exp(-\eta x)$, known as the exponential distribution. Henceforth, it can be stated that when the correlation among individual firing of neurons is low enough, the power distribution of an electrode is exponential. Any deviation from this distribution might suggest the presence of a correlation between the firing of neurons.
  Scaling the variable $x$ as $y=ln(x),$ the scaled distribution 
  \begin{equation}
    P_Y(y)= \left|\frac{dx}{dy}\right| \:\: P_Y(y)\:=\eta \: exp(y-exp(\eta\: y)).
    \label{eqn:unimodal}
  \end{equation}
  
A deviation from exponential statistics indicates the presence of temporal correlation. Moreover, in some cases, there exists a switching between the two states which can result in bimodal exponential statistics given by,
\begin{equation}
    P_{XX}(x)=(1-\delta)\:\eta_1 \: exp(-\eta_1 x) +\delta\:\eta_2 \: exp(-\eta_2 x).
    \label{eqn:bimodal}
\end{equation}
We expect the power distribution to follow an exponential distribution due to its stochastic nature. We find the parameters for the best fit (both unimodal (Eq.~\ref{eqn:unimodal}) and bimodal (Eq.~\ref{eqn:bimodal})) of the empirical power distributions at various time windows using maximum likelihood estimation, and use the Bayesian information criterion (BIC) to compare the unimodal and the bimodal exponential distributions. BIC incorporates a penalty term proportional to the number of parameters used while fitting, and is defined as
$BIC=-2 \times ln(L)-ln(n)\times number\: of \:parameters$.
where $L$ is the likelihood and $n$ is the number of bins used for the distribution. The penalty term is the deciding factor in cases where both models fit the data equally. The lower the BIC value, the better is the fit. We calculate $\Delta BIC$ which is defined as $BIC(unimodal)-BIC(bimodal)$. Therefore, a negative $BIC$ value infers a better unimodal fit and the vice versa.
\begin{figure}
 \includegraphics[width=0.99\linewidth]{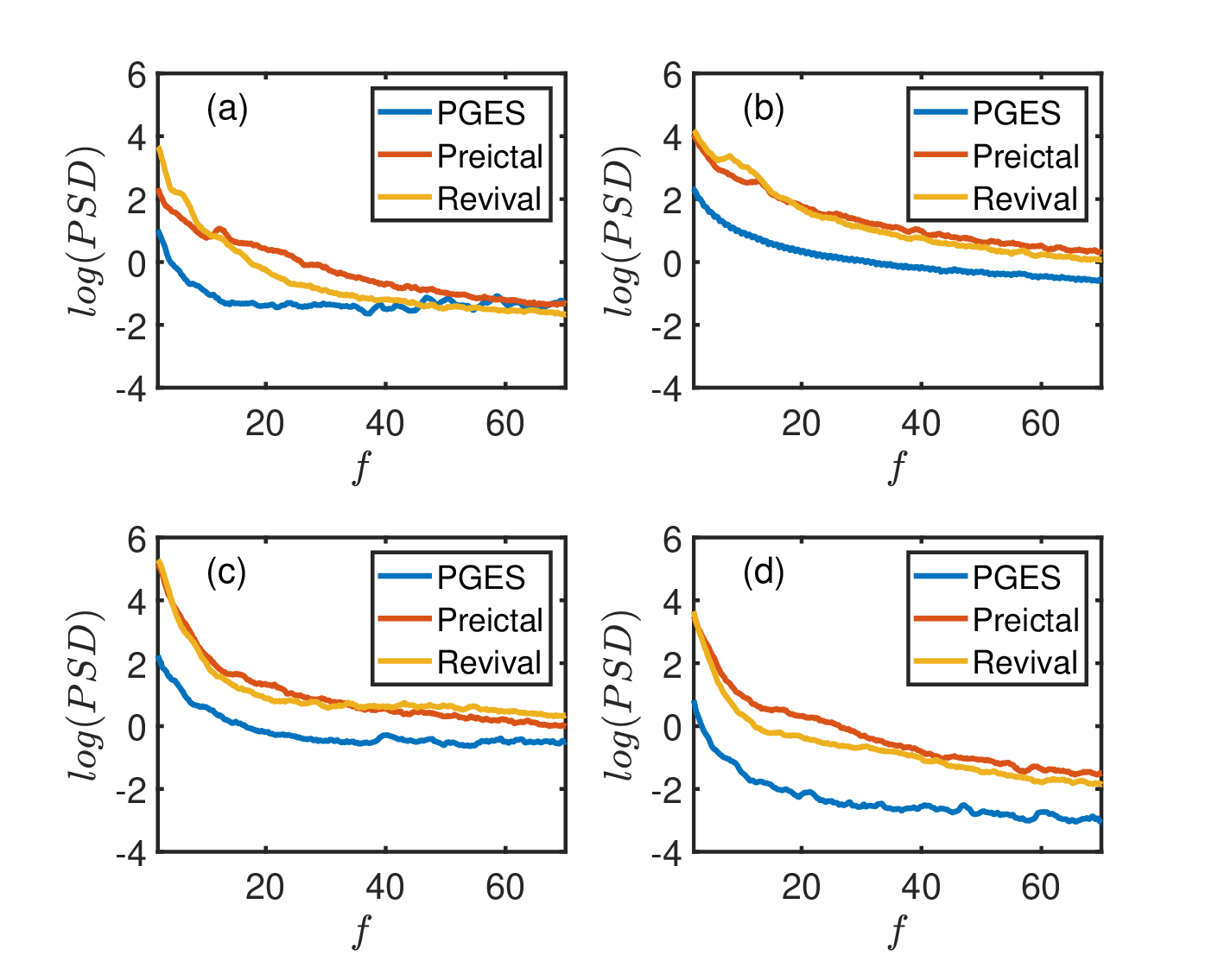}
 \caption{$log(PSD)$ vs $f$ : Power spectrum for (a) Subject 1, (b) Subject 2, (c) Subject 3, and (d) Subject 4. The blue, red and yellow lines represent the PGES, preictal region and revival state, respectively.}     
 \label{fig:pow_spec} 
\end{figure}

\begin{figure*}[t!]
    \includegraphics[width=0.99\linewidth]{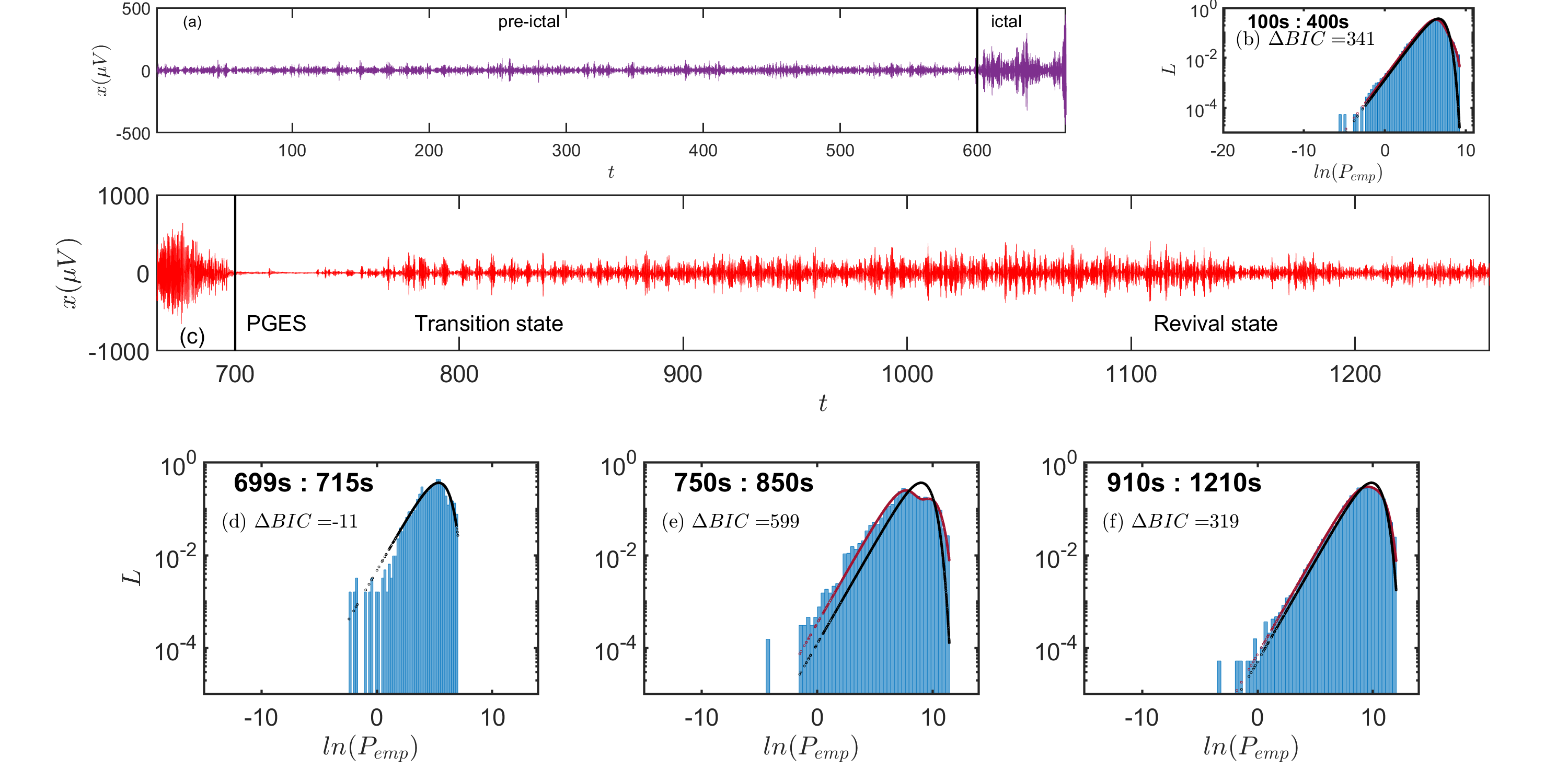}
      \caption{Real-valued EEG time series ($x$ vs. $t$) and corresponding power distributions over various time intervals for subject 1 (Frontal region) in the $\delta$ band. (a) Time series for the preictal and ictal stages, separated by a solid black line; (b) Power distribution for the preictal region from 100s to 400s; (c) Time series for the postictal stage; (d) Power distribution for the PGES (suppressed state) in the time interval from 699s to 715s; (e) Power distribution for the transition state between 750s and 850s; (f) Power distribution for the revival state from 910s to 1210s. The solid black and red lines  corresponds to the unimodal and bimodal exponential fit, respectively.} \label{fig:timeseries_delta}
\end{figure*}

\subparagraph{{Model.}}
The objective here is to identify a suitable dynamical model to explain the bifurcations or phase transitions that occur in the postictal region of a tonic-clonic seizure. Two major models exhibit a bistable region shared between a limit cycle and a fixed point within a region of their parameter space; saddle-homoclinic bifurcation and subcritical Hopf bifurcation. In saddle-homoclinic bifurcation, a pair of saddle points emerges in the phase space where a limit cycle already exists. As the bifurcation parameter increases, the bistable region in phase space disappears due to the collision of a stable fixed point with a stable limit cycle. In subcritical Hopf bifurcation, a stable fixed point bifurcates into an unstable fixed point and an unstable limit cycle. This unstable limit cycle later changes its stability, leading to a hysteresis effect (Fig.~\ref{fig:bif}). However, in saddle-homoclinic bifurcation, the fixed point is typically parameter dependent, which is analogous to a changing baseline in EEG data \cite{jirsa2014nature}. Since the system considered here does not exhibit any significant baseline shift, we proceed with the "subcritical Hopf model", described by the following differential equation,
\begin{equation}
    \dot r=\lambda r^3+\beta r -r^5 \:, \;\; \dot \theta=\omega.
\label{eqn:sub_hopf}
\end{equation}
This system of differential equations can be written in the cartesian coordinates as,
\begin{equation}
    \dot x=(\lambda r^2+\beta  -r^4)x-\omega y \:, \;\; 
     \dot y=(\lambda r^2+\beta  -r^4)y+\omega x
\label{eqn:sub_hopf_cart}\nonumber
\end{equation}
Here, $\omega$ is the intrinsic frequency, and $\beta$ and $\lambda$ are model parameters. The fixed points of these equations are given by $r^*_{\pm}=\sqrt{\frac{\lambda\pm \sqrt{\lambda^2+4\beta}}{2}}$, where $r^*_+$ is the stable branch and $r^*_-$ is the unstable branch. The parameter $\lambda$ controls the forward critical point of the limit cycle, and the fixed point always loses its stability at $\beta=0$. Therefore, the parameter $\lambda$ actually controls the width of the hysteresis. The parameter $\beta$ is the bifurcation parameter required to obtain sub critical Hopf bifurcation. We tune these parameters to switch between the various states of our interest. However, in this model, there is a bistable state which is dependent on the initial conditions. Hence, to obtain the switching between the states we add additive and multiplicative Gaussian white noise to the system (Eq.~\ref{eqn:sub_hopf}) \cite{freyer2012canonical}.
\begin{equation}
    \dot r=\lambda r^3+\beta r -r^5 + D_{add} \xi(t) + D_{mull} r \zeta(t)\:, \;\;
\label{eqn:sub_hopf_noise}
\end{equation}
Here, $\xi(t)$ and $\zeta(t)$ represent Gaussian white noise with zero mean and unit variance, respectively, and $D_{mull}$ and $D_{add}$ correspond to the additive and multiplicative noise strengths, respectively. We calculate the amplitude as $A(t)=\sqrt{x(t)^2+y(t)^2}$ and the power at time $t$ as $P_{model}(t)=A(t)^2$. We expect the distribution of globally stable states to have a unimodal exponential distribution, and bistable state to have bimodal exponential distribution at appropriate values of $\eta$. 
The dynamics of this model can be understood through the bifurcation plot (Fig.~\ref{fig:bif}). For, $\lambda=4$ and $\omega=3$ sub-critical Hopf bifurcation takes place at $\beta=0$. For $\beta>0$ we have a globally stable limit cycle with its amplitude scaling as $\sqrt{\beta}$. Moreover, we encounter a stable fixed point for  $\beta<-4.0$ which again is globally stable. However, for $-4<\beta<0$ there is a bistable region where the basin of attraction is shared by the stable fixed point and the stable limit cycle. 

The range of $x(t)$ varies from one patient to the other, and the same is true for the various channels in each patient. However, in our model, the range depends on the applied noise strengths $D_{add}$ and $D_{mull}$. Numerical simulations are only possible for a certain noise strength before the solution diverges. We transform Eq.~\ref{eqn:sub_hopf_noise} using a scale factor to counteract this. This scaling preserves the bifurcation and other dynamical properties of the system. Substituting $r\rightarrow r/s$ we get,
\begin{equation}
    \dot r=(\lambda \frac{r^2}{s^2}+\beta-\frac{r^4}{s^4})r + s D_{add} \xi(t) + D_{mull} r \zeta(t)\:,\;\;
\label{eqn:sub_hopf_noise_scaled}
\end{equation}
where $s$ is the scale factor.
Eq.~\ref{eqn:sub_hopf_noise_scaled} was simulated using the Heun's method with step size $dt=0.001$. All simulations were performed for $10^6$ steps, and the initial $5\times 10^5$ steps were discarded as transient. 

\subparagraph{\bf{Determining model parameter values: }} Here, our objective is to find the model parameters of Eq.~\ref{eqn:sub_hopf_noise_scaled} for which it will produce a power-law distribution statistically similar to the empirical power-law distribution. We perform non-parametric two-sample statistical tests like the Mann-Whitney test, mean-based permutation test, and Kolmogorov-Smirnov test to assess the fact that the differences between the two samples are not statistically significant.

\textit{Mann-Whittney test: }The two-sample Mann-Whitney test evaluates whether there exists a statistically significant difference in the distribution of two independent groups.  The test is based on ranking all the observations from both groups together and then comparing the ranks between the two groups. By calculating the sum of the ranks for each group, the Mann-Whitney test assesses whether one group tends to have higher or lower ranks than the other. The resulting U-statistic, which reflects the difference between the rank sums, is then used to determine the significance of the observed difference. A p-value is obtained by comparing the observed U-statistic and its distribution under the null hypothesis.

\textit{Mean based permutation test: } The null hypothesis states that the difference of the mean between the two distributions is not statistically significant. The data from both groups are combined into a single pooled dataset, which is then randomly shuffled to create a permuted dataset. After each shuffle, the observations are reassigned to the two groups, and the difference in their means is calculated. This process is repeated 10,000 times. The $p$-value is determined by calculating the proportion of permuted mean differences that are as extreme or more extreme than the observed mean difference of the original data.

\textit{Kolmogorov-Smirnov test:} The two sample Kolmogorov-Smirnov (KS) test is a non parametric statistical test used to determine whether two independent samples are drawn from the same continuous distribution. It compares the empirical cumulative distribution functions of the two samples and calculates the maximum absolute difference between them which determines the $p$-value. This approach is particularly useful for identifying distributional shifts or differences in the shape of the distributions.

For all tests, a significance level of 0.05 was used to determine whether the null hypothesis is true. The three tests used in this study are based on different criteria for comparing distributions. The Mann-Whitney test focuses on the median, the mean-based permutation test on the mean, and the KS test compares the cumulative distributions. This diverse statistical approach allows us to assess the similarity of distributions from multiple perspectives, thereby, enhancing the validity of our hypothesis. However, due to the stochastic nature of Eq.~\ref{eqn:sub_hopf_noise_scaled}, a particular set of parameters can yield different power distributions. To account for this variability, we simulate 100 power distributions for each set of parameters and report the fraction of simulations where the null hypothesis was true.
\begin{figure}
 \includegraphics[width=0.99\linewidth]{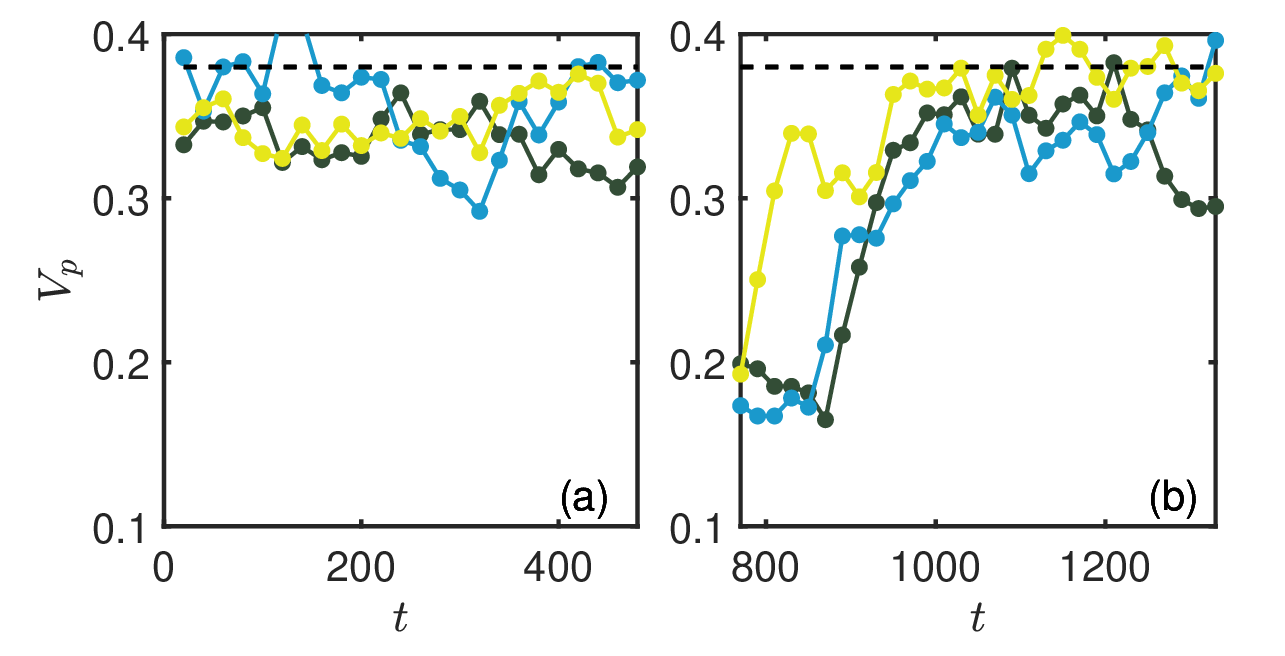}\\
 \caption{$V_p$ vs $t$ (Subject 1). (a) The peak value of the preictal power distribution remains constant over time, indicating the absence of a phase transition in this region.
 (b) The peak value of the postictal power distribution is slightly lower than $\frac{1}{e}$, suggesting a bimodal distribution. Blue, black and yellow circle corresponds to Orbitofrontal, Posterior hippocampus and Thalamus region, respectively. }   
 \label{fig:peak} 
\end{figure}
\begin{figure}[t!]
    \includegraphics[width=0.99\linewidth]{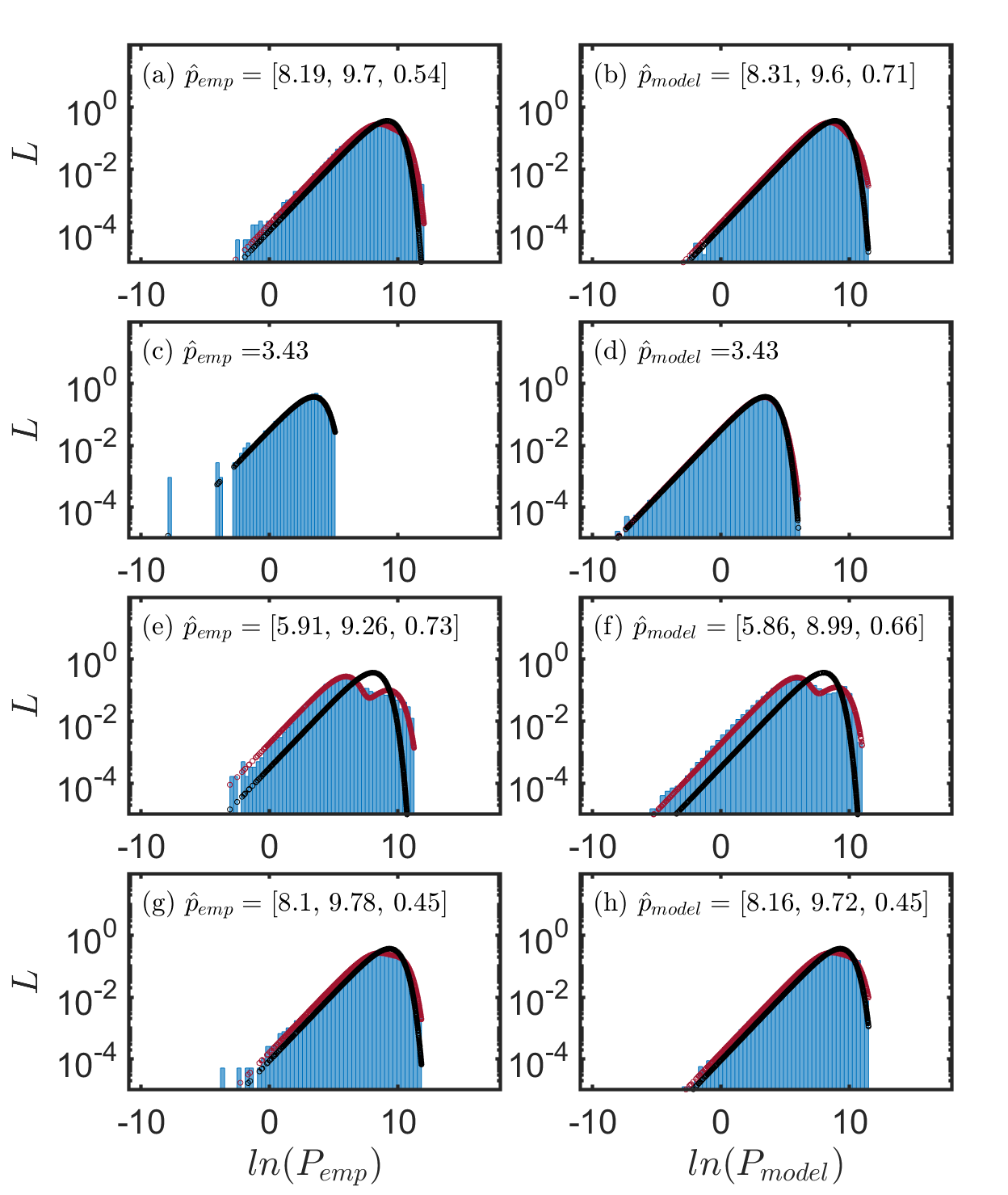}
      \caption{$L$ vs. $\ln(P_{emp})$ ($\ln(P_{model})$) for Subject 4. The power $P_{model}$ is calculated using Eq.~\ref{eqn:sub_hopf_noise_scaled}. (a) Distribution of $P_{emp}$ in the preictal state, (b) distribution of $P_{model}$ at $\Delta\beta=0.99$, $\lambda=4.0$, $D_{add}=19$, $D_{mull}=0$, $2\ln s=11.6$, (c) distribution of $P_{emp}$ in the PGES state, (d) distribution of $P_{model}$ at $\Delta\beta=15$, $\lambda=4.0$, $D_{add}=15$, $D_{mull}=0$, $2\ln s=6.8$, (e) distribution of $P_{emp}$ in the transition state, (f) distribution of $P_{model}$ at $\Delta\beta=0.4$, $\lambda=8.0$, $D_{add}=15$, $D_{mull}=70$, $2\ln s=7.6$, (g) distribution of $P_{emp}$ in the revival state, (h) distribution of $P_{model}$ at $\Delta\beta=0.85$, $\lambda=4.0$, $D_{add}=15$, $D_{mull}=0$, $2\ln s=9.9$. $\Delta \beta=\frac{\beta}{-4}$.}
 \label{fig:model_hist}
\end{figure}

\begin{figure*}[t!]
 \includegraphics[width=0.99\linewidth]{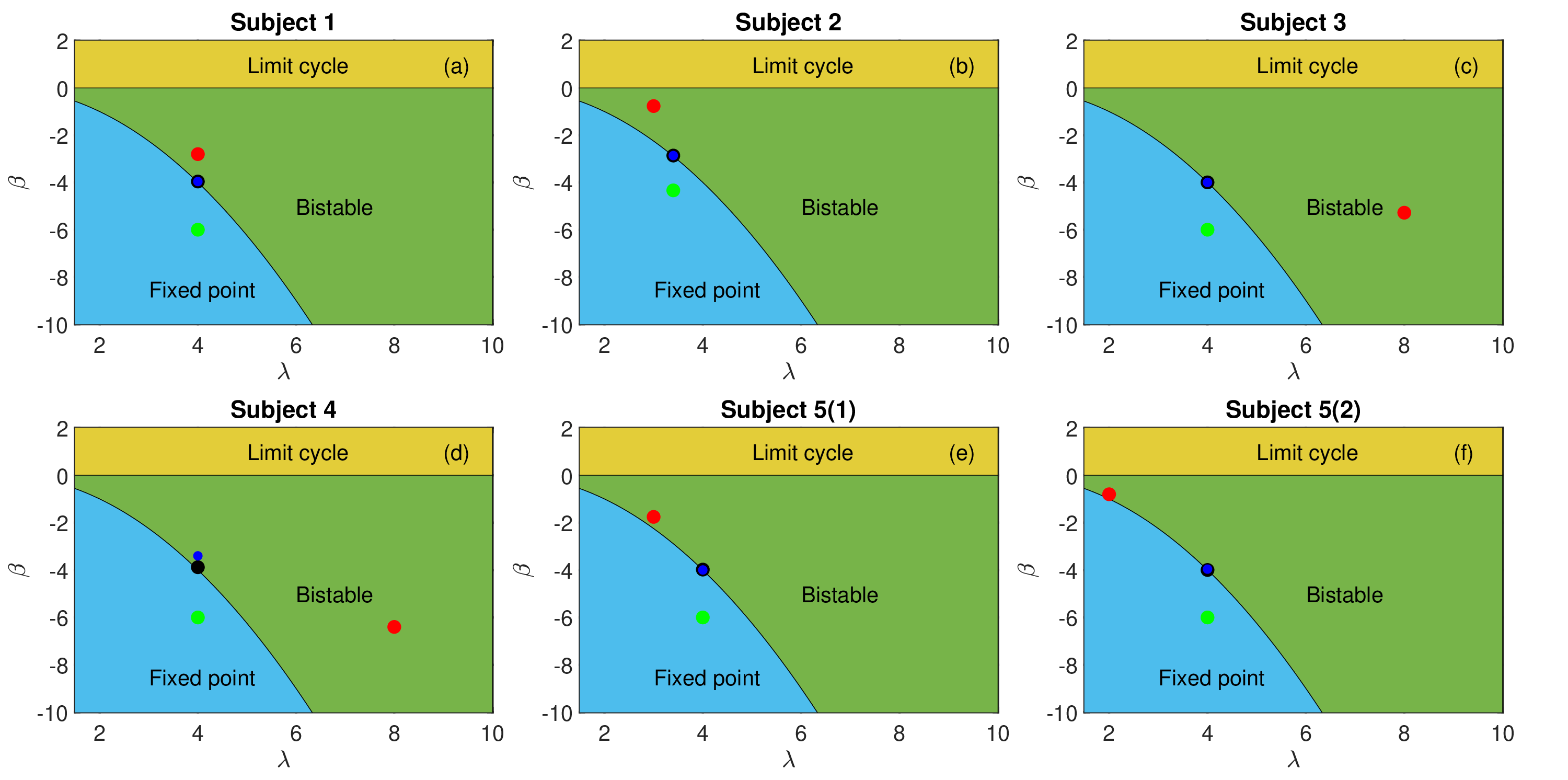}
 \caption{$\beta$ vs $\lambda$ bifurcation plot for normal form of subcritical Hopf given by Eq.~ \ref{eqn:sub_hopf_noise} with $D_{add}=0$ and $D_{mull}=0$. black solid circle corresponds to preictal state, blue solid circle corresponds to revival state, red solid circle corresponds to transition state, green solid circle corresponds to PGES state. Each sub figure corresponds to a seizure, (a) subject 1, (b) subject 2, (c) subject 3, (d) subject 4, (e) subject  5(1), (f) subject 5 (2). } 
 \label{fig:phase_space} 
\end{figure*}

\paragraph{\bf {Results:}}
Generally, detection of PGES from EEG data analysis is an intensive and manual process \cite{mier2020categorisation}. However, identification of PGES can also be done by analyzing the time-series data and the power spectrum \cite{marchi2019postictal,tao2013tonic,bateman2019postictal}. In $\delta$ and $\theta$ frequency bands, the power spectral density($PSD$) is much lower than the preictal counterpart. The subjects reported in Fig.~\ref{fig:pow_spec} illustrates that the difference in the $PSD$ for lower frequency bands ($\delta, \theta$) is considerably higher than the high frequency bands ($\alpha, \beta, \gamma$) (Fig. ~\ref{fig:pow_spec}). This indicates reduced brain activity immediately after a seizure compared to the preictal state. According to the definition of PGES, the absence of activity below $10\mu V$ for at least 1 second within 30 seconds of seizure can be classified as PGES. Therefore, combining these observations can serve as a reliable marker for identifying PGES.

Upon analyzing the empirical power ($P_{emp}$) distribution in the preictal region, we observe a bimodal exponential distribution in all the subjects (Fig.~\ref{fig:timeseries_delta}(a,b)).  We find positive $\Delta BIC$ values for the subjects with bimodal distributions (Supplementary material \cite{SM}). Analysis of $\delta$ and $\theta$ bands reveals three stages of the postictal state: PGES, the transition state and the revival state. In the first stage, the EEG signal is wholly attenuated immediately after the seizure. This period is referred to as the PGES (Fig.~\ref{fig:timeseries_delta}(c,d)), which lasts approximately 15 to 20 seconds in our subjects. During PGES, we report an unimodal exponential distribution with negative values of $\Delta BIC$ in all subjects with the average $\Delta BIC$ around $-10.74$. The resurgence of the background activity marks the end of PGES, typically intermittent slow activity. This phase in the time series is characterized by a mixture of signal bursts and suppression (Fig.~\ref{fig:timeseries_delta}(c)). We refer to this phase as the transition region, representing an intermediate state between PGES and normal brain function. In this region, intervals of suppressed signals appear as a low-power mode. At the same time, sudden bursts correspond to a high-power mode in the power distribution of the transition region (Fig.~\ref{fig:timeseries_delta}(e)). This pattern suggests the presence of two distinct states, with the system switching between them. The $\Delta BIC$ values in this state are positive and consistently higher for all subjects compared to other time-series intervals (preictal and revival), indicating stronger bimodality. The $\langle \Delta BIC \rangle$ over all subjects in transition state is $662$, which is much higher than revival ($\langle \Delta BIC \rangle=3 08$) and preictal ($\langle \Delta BIC \rangle = 203$) states. This further supports the idea of two coexisting states with transitions between them. As the series progresses, these fluctuations become more frequent, resembling the preictal state. Finally, the time series transitions into a state similar to the preictal phase, which we term the revival state (Fig.~\ref{fig:timeseries_delta}(c,f)). The similarity between the preictal and revival states is evident from their $\Delta BIC$ values, indicating that the brain has returned to normal function. Moreover, the power spectral density (Fig.~\ref{fig:pow_spec}) of preictal and revival states are similar unlike the PGES, further asserting the similarity between these two states.

To understand the transition in the postictal state, we analyze the temporal behaviour of the peak value $V_p$ (highest frequency) in the power distribution. The normalized unimodal exponential distribution has a consistent $V_p$ of $1/e$ at $P=\frac{1}{\eta}$, regardless of the distribution's shape parameter (Eq.~\ref{eqn:unimodal}). However, in the case of a normalized bimodal exponential distribution, $V_p$ depends on the shape and proportionality parameters and is always equal to or less than $1/e$, the maximum peak value (Eq.~\ref{eqn:bimodal}). A more significant deviation from this maximum peak value indicates increased bimodality in the empirical distribution. Fig.~\ref{fig:peak}(a) shows that $V_p$ for the preictal power distribution remains constant with time, indicating no phase transition in this region. Furthermore, for most subjects, $V_p$ is slightly lower than the maximum, suggesting a mild degree of bimodality.

Fig.~\ref{fig:timeseries_delta}(d) further illustrates that in the suppressed region, which is unimodal, the peak value \( V_p \) remains at \( 1/e = 0.37 \). In Fig.~\ref{fig:peak}(b), we start from the transition state and continue up to the revival state. Here, we observe a sudden drop in \( V_p \) (transition state), followed by a gradual increase over time in the postictal region, as it progresses toward the revival state (Fig.~\ref{fig:peak}(b)). This confirms the earlier observation of a transition from a state with higher bimodality (more positive \( \Delta BIC \) values and lower \( V_p \)) to a state with lower bimodality following the suppressed state.  Interestingly, Fig.~\ref{fig:peak} also shows that all channels exhibit varying degrees of bimodality and different durations of the transition state. In the case of Subject 1, we observe that the Frontal region exhibits a more prominent transition state.

Next, we demonstrate that the dynamical model accurately simulates the four stages: preictal state, postictal suppression state, transition state, and revival state (Fig.~\ref{fig:model_hist}). Stages exhibiting bimodality fall within the bistable region of the dynamical system, while those showing an unimodal state likely correspond to the fixed-point region of the system. Our analysis shows that the dynamical model best mimics the preictal region when the parameters are on the verge of the bistable region and the fixed point, with a tendency toward the bistable region for all subjects Fig.~\ref{fig:phase_space}(a-f). Furthermore, Fig.~\ref{fig:model_hist} reports the parameter values of the fitted exponential distribution (unimodal or bimodal) for both the empirical and model generated power distributions, denoted as $\hat{p}_{emp}$ and $\hat{p}_{model}$, respectively. For an unimodal distribution, the parameters are given by,  
$
\hat{p}_{emp(model)} = \frac{1}{\eta}.
$  
For a bimodal distribution, they are defined as:  
$
\hat{p}_{emp(model)} = \left[\frac{1}{\eta_1}, \frac{1}{\eta_2}, \delta\right]
$ ,where \(\eta, \eta_1, \eta_2\), and \(\delta\) have the same definitions as in Eqs.~\ref{eqn:unimodal} and \ref{eqn:bimodal}. From the Fig.~\ref{fig:model_hist}, it is evident that the parameter values of the empirical data closely match those of the model generated data for each state.  

The PGES state is characterized by parameters in the fixed-point region for all subjects. The scale factor (\(s\)), which is one of the factors controlling \(\eta\), is lower in this state compared to the preictal and revival counterparts. A lower scale factor indicates a reduced mean in the unimodal distribution, which suggests a suppression of signal fluctuations. This highlights an important observation: PGES is associated with reduced variability in brain activity. Additionally, during the transition phase, the parameter set for all subjects falls within the bistable region Fig.~\ref{fig:phase_space}. The value of ($\beta$) depends on the position and proportion of the two modes in the bimodal distribution. A higher proportion of the high-power mode results in a ($\beta$) value deeper within the bistable region, further away from the boundary between the fixed-point and bistable regions in phase space. Finally, the parameters in the revival region closely resemble those of the preictal state. The transition from the suppressed state to the transition state can be interpreted as a shift from a fixed point to a bistable state in the model. However, since multiple parameters must be adjusted to progress from PGES to the transition state and the revival state, we refrain from classifying this as a strict bifurcation and instead call it a phase transition.  

As mentioned earlier, we performed three statistical tests, namely, the Mann-Whitney test, the mean-based permutation test, and the Kolmogorov-Smirnov test to assess whether the null hypothesis (that the empirical and simulated power distributions come from the same distribution) can be rejected. While, we successfully identified model parameters where the null hypothesis was supported by all the three tests for some of the PGES and transition states, the Kolmogorov-Smirnov test often failed for highly bimodal distributions like the transition state. This issue primarily stems from the nature of the test, which compares the cumulative distributions of the samples, making it sensitive to small fluctuations that can lead to the  failure of the test \cite{SM}.

PGES is a scalp EEG phenomenon. A similar phenomenon, called intercranial postictal attenuation (IPA), has been observed in intracranial EEG exhibiting similar EEG signal suppression following a seizure, much like PGES. The primary difference between them is the $\gamma$ wave activity, which is seen in IPA while being absent in PGES. Additionally, IPA has been reported to show a mixture of high-amplitude and low-amplitude segments in the time series, resembling the patterns observed in our system \cite{marchi2019postictal,bateman2019postictal}. Although it is unknown whether PGES and IPA are manifestations of the same phenomenon, we point out another similarity (apart from those reported earlier \cite{bateman2019postictal}) found between the two phenomena i.e. existence of the bistable transition state. 

\paragraph{\bf{Conclusion:}}
Postictal dynamics vary across brain regions and individuals but typically begin with a suppressed phase marked by significant signal attenuation. This is followed by a transitional phase where bursts of high-amplitude activity intermittently interrupt the suppressed state, eventually leading to recovery characterized by a return to preictal-like activity. In this study, we analyzed the power distribution across EEG channels and found that the suppressed state exhibits a unimodal distribution, while the transitional phase displays pronounced bimodality. To capture this behavior, we employed the subcritical Hopf normal form, identifying parameter regimes that reproduce these empirical features. While the complexity of parameter interactions prevents us from definitively attributing the observed transitions to a classical subcritical Hopf bifurcation, our results support a transition from a fixed-point regime to a bistable state as a plausible underlying mechanism of EEG recovery during PGES.

A natural extension of this work involves incorporating coupling into the current model to better replicate EEG dynamics across different pathological states. Exploring the influence of various coupling schemes may yield insights into the mechanisms governing inter-regional coordination. Furthermore, integrating time-varying functional connectivity into dynamical modeling could enhance the accuracy and interpretability of models describing neurobiological phenomena.

\begin{acknowledgments}
    SJ gratefully acknowledges SERB Power grant SPF/2021/000136, and VAJRA project VJR/2019/000034. The authors acknowledges the assistance of OpenAI for grammatical corrections, punctuation improvements, and rephrasing certain sentences to enhance clarity and readability. All technical content, ideas, and interpretations presented in this work are solely those of the authors.
\end{acknowledgments}


\begin{thebibliography}{50}
\bibliographystyle{apsrev4-2}

\bibitem{milton1989complex} Milton, J. G., Longtin, A., Beuter, A., Mackey, M. C., \& Glass, L. Complex dynamics and bifurcations in neurology. {\em Journal of Theoretical Biology} {\bf 138(2)}, 129–147 (1989).

\bibitem{ashwin2016mathematical} Ashwin, P., Coombes, S., \& Nicks, R. Mathematical frameworks for oscillatory network dynamics in neuroscience. {\em The Journal of Mathematical Neuroscience} {\bf 6}, 1–92 (2016).
\bibitem{ren2021connectivity} Ren, X., Brodovskaya, A., Hudson, J. L., \& Kapur, J. Connectivity and neuronal synchrony during seizures. {\em Journal of Neuroscience} {\bf 41(36)}, 7623–7635 (2021).

\bibitem{truccolo2014neuronal} Truccolo, W., Ahmed, O. J., Harrison, M. T., Eskandar, E. N., Cosgrove, G. R., Madsen, J. R., Blum, A. S., Potter, N. S., Hochberg, L. R., \& Cash, S. S. Neuronal ensemble synchrony during human focal seizures. {\em Journal of Neuroscience} {\bf 34(30)}, 9927–9944 (2014).


\bibitem{hollister1980hormones} Hollister, L. E., Davis, K. L., \& Davis, B. M. Hormones in the treatment of psychiatric disorders. In D. T. Kreiger \& J. C. Hughes (Eds.), {\em Neuroendocrinology} (pp. 167–175). Sinauer (1980).





\bibitem{uhlhaas2006neural} Uhlhaas, P. J., \& Singer, W. Neural synchrony in brain disorders: Relevance for cognitive dysfunctions and pathophysiology. {\em Neuron} {\bf 52(1)}, 155–168 (2006).

\bibitem{aron2016neural} Aron, L., \& Yankner, B. A. Neural synchronization in Alzheimer's disease. {\em Nature} {\bf 540(7632)}, 207–208 (2016).

\bibitem{moosavi2022critical} Moosavi, S. A., Jirsa, V. K., \& Truccolo, W. Critical dynamics in the spread of focal epileptic seizures: Network connectivity, neural excitability, and phase transitions. {\em PLOS One} {\bf 17(8)}, e0272902 (2022).

\bibitem{lhatoo2010electroclinical} Lhatoo, S. D., Faulkner, H. J., Dembny, K., Trippick, K., Johnson, C., \& Bird, J. M. An electroclinical case-control study of sudden unexpected death in epilepsy. {\em Annals of Neurology} {\bf 68(6)}, 787–796 (2010).

\bibitem{rajakulendran2015postictal} Rajakulendran, S., \& Nashef, L. Postictal generalized EEG suppression and SUDEP: A review. {\em Journal of Clinical Neurophysiology} {\bf 32(1)}, 14–20 (2015).
\bibitem{surges2011postictal} Surges, R., Strzelczyk, A., Scott, C. A., Walker, M. C., \& Sander, J. W. Postictal generalized electroencephalographic suppression is associated with generalized seizures. {\em Epilepsy \& Behavior} {\bf 21(3)}, 271–274 (2011).
\bibitem{xu2016postictal} Xu, J., Jin, B., Yan, J., Wang, J., Hu, J., Wang, Z., Chen, Z., Ding, M., Chen, S., \& Wang, S. Postictal generalized EEG suppression after generalized convulsive seizures: A double-edged sword. {\em Clinical Neurophysiology} {\bf 127(4)}, 2078–2084 (2016).

\bibitem{marchi2019postictal} Marchi, A., Giusiano, B., King, M., Lagarde, S., Trébuchon-Dafonseca, A., Bernard, C., Rheims, S., Bartolomei, F., \& McGonigal, A. Postictal electroencephalographic (EEG) suppression: A stereo-EEG study of 100 focal to bilateral tonic–clonic seizures. {\em Epilepsia} {\bf 60(1)}, 63–73 (2019).

\bibitem{tao2013tonic} Tao, J. X., Yung, I., Lee, A., Rose, S., Jacobsen, J., \& Ebersole, J. S. Tonic phase of a generalized convulsive seizure is an independent predictor of postictal generalized EEG suppression. {\em Epilepsia} {\bf 54(5)}, 858–865 (2013).




\bibitem{mier2020categorisation} Mier, J. C., Kim, Y., Jiang, X., Zhang, G.-Q., \& Lhatoo, S. Categorisation of EEG suppression using enhanced feature extraction for SUDEP risk assessment. {\em BMC Medical Informatics and Decision Making} {\bf 20}, 1–6 (2020).

\bibitem{liu2023task} 
Liu, Z., Han, F., \& Wang, Q. Task-relevant brain dynamics among cognitive subsystems induced by regional stimulation in a whole-brain computational model. {\em Physical Review E}, {\bf 108}(4), 044402 (2023). https://doi.org/10.1103/PhysRevE.108.044402


\bibitem{cabral2012modeling} Cabral, J., Hugues, E., Kringelbach, M. L., \& Deco, G. Modeling the outcome of structural disconnection on resting-state functional connectivity. {\em NeuroImage} {\bf 62(3)}, 1342–1353 (2012).

\bibitem{shine2015estimation} Shine, J. M., Koyejo, O., Bell, P. T., Gorgolewski, K. J., Gilat, M., \& Poldrack, R. A. Estimation of dynamic functional connectivity using Multiplication of Temporal Derivatives. {\em NeuroImage} {\bf 122}, 399–407 (2015).

\bibitem{hutchison2013dynamic} Hutchison, R. M., Womelsdorf, T., Allen, E. A., Bandettini, P. A., Calhoun, V. D., Corbetta, M., Della Penna, S., Duyn, J. H., Glover, G. H., Gonzalez-Castillo, J., Handwerker, D. A., Keilholz, S., Kiviniemi, V., Leopold, D. A., de Pasquale, F., Sporns, O., Walter, M., \& Chang, C. Dynamic functional connectivity: Promise, issues, and interpretations. {\em NeuroImage} {\bf 80}, 360–378 (2013).

\bibitem{cabral2017functional} Cabral, J., Kringelbach, M. L., \& Deco, G. Functional connectivity dynamically evolves on multiple time-scales over a static structural connectome: Models and mechanisms. {\em NeuroImage} {\bf 160}, 84–96 (2017).

\bibitem{deco2012ongoing} Deco, G., \& Jirsa, V. K. Ongoing cortical activity at rest: Criticality, multistability, and ghost attractors. {\em Journal of Neuroscience} {\bf 32(10)}, 3366–3375 (2012).

\bibitem{freyer2009bistability} Freyer, F., Aquino, K., Robinson, P. A., Ritter, P., \& Breakspear, M. Bistability and non-Gaussian fluctuations in spontaneous cortical activity. {\em Journal of Neuroscience} {\bf 29(26)}, 8512–8524 (2009).


\bibitem{deco2017single} Deco, G., Cabral, J., Woolrich, M. W., Stevner, A. B. A., Van Hartevelt, T. J., \& Kringelbach, M. L. Single or multiple frequency generators in on-going brain activity: A mechanistic whole-brain model of empirical MEG data. {\em NeuroImage} {\bf 152}, 538–550 (2017).

\bibitem{ghosh2008cortical} Ghosh, A., Rho, Y., McIntosh, A. R., Kötter, R., \& Jirsa, V. K. Cortical network dynamics with time delays reveals functional connectivity in the resting brain. {\em Cognitive Neurodynamics} {\bf 2(2)}, 115–120 (2008).


\bibitem{freyer2012canonical} Freyer, F., Roberts, J. A., Ritter, P., \& Breakspear, M. A canonical model of multistability and scale-invariance in biological systems. {\em PLOS ONE} (2012).


\bibitem{nazarimehr2018does} Nazarimehr, F., Hashemi Golpayegani, S. M. R., \& Hatef, B. Does the onset of epileptic seizure start from a bifurcation point?. {\em The European Physical Journal Special Topics} {\bf 227}, 697–705 (2018).

\bibitem{breakspear2006unifying} Breakspear, M., Roberts, J. A., Terry, J. R., Rodrigues, S., Mahant, N., \& Robinson, P. A. A unifying explanation of primary generalized seizures through nonlinear brain modeling and bifurcation analysis. {\em Cerebral Cortex} {\bf 16(9)}, 1296–1313 (2006).

\bibitem{jirsa2014nature} Jirsa, V. K., Stacey, W. C., Quilichini, P. P., Ivanov, A. I., \& Bernard, C. On the nature of seizure dynamics. {\em Brain} {\bf 137(8)}, 2210–2230 (2014).


\bibitem{dutta1} Dutta, S., Alamoudi, O., Vakilna, Y. S., Pati, S., \& Jalan, S. Oscillation quenching in Stuart-Landau oscillators via dissimilar repulsive coupling. {\em Physical Review Research} {\bf 5(1)}, 013074 (2023).

\bibitem{verma} Verma, U. K., Sharma, A., Kamal, N. K., \& Shrimali, M. D. First order transition to oscillation death through an environment. {\em Physics Letters A} {\bf 382(32)}, 2122–2126 (2018).

\bibitem{cohen_book} Cohen, M. X. Analyzing neural time series data: theory and practice. {\em MIT press}. (2014)

\bibitem{feller} Feller, W. An introduction to probability theory and its applications, Volume 1. {\em John Wiley \& Sons}. (1991)

\bibitem{gonen2012techniques} Gonen, F. F., \& Tcheslavski, G. V. Techniques to assess stationarity and gaussianity of EEG: An overview. {\em International Journal of Bioautomation} {\bf 16(2)}, 135 (2012).

\bibitem{lion1953method} Lion, K. S., \& Winter, D. F. A method for the discrimination between signal and random noise of electrobiological potentials. {\em Electroencephalography and Clinical Neurophysiology} {\bf 5(1)}, 109–111 (1953).

\bibitem{saunders1963amplitude} Saunders, M. G. Amplitude probability density studies on alpha and alpha-like patterns. {\em Electroencephalography and Clinical Neurophysiology} {\bf 15(5)}, 761–767 (1963).

\bibitem{mcewen1975modeling} McEwen, J. A., \& Anderson, G. B. Modeling the stationarity and gaussianity of spontaneous electroencephalographic activity. {\em IEEE Transactions on Biomedical Engineering} {\bf (5)}, 361–369 (1975).

\bibitem{bateman2019postictal} Bateman, L. M., Mendiratta, A., Liou, J.-Y., Smith, E. J., Bazil, C. W., Choi, H., McKhann, G. M., Pack, A., Srinivasan, S., \& Schevon, C. A. Postictal clinical and electroencephalographic activity following intracranially recorded bilateral tonic–clonic seizures. {\em Epilepsia} {\bf 60(1)}, 74–84 (2019).

\bibitem{SM} Supplementary material contains  the BIC values and distribution parameters for empirical data and model generated data along with results of statistical test comparing them.

\end{thebibliography}
\end{document}